**Climate Sensitivity Derived from First-Principles Temperature Feedback Circuits**

Ming Cai[1,*,†], Jie Sun[1,†], Zhe-Min Tan[2], Xiaoming Hu[3], Guosheng Liu[1], Zhaohua Wu[1]

[1]Department of Earth, Ocean, and Atmospheric Science, Florida State University, Tallahassee, FL, 32306, USA.

[2]School of Atmospheric Sciences, and State Key Laboratory of Severe Weather Meteorological Science and Technology, NJU, Nanjing University, Nanjing, China.

[3]School of Atmospheric Sciences, Sun Yat-sen University and Southern Marine Science and Engineering Guangdong Laboratory (Zhuhai), Zhuhai, P. R. China.

*Corresponding author(s). E-mail(s): mcai@fsu.edu;

†These authors contributed equally to this work.

**Abstract**

A long-standing challenge in climate science since the Charney Report in 1979 is to determine how much the Earth's climate warms in response to a unit external energy input at the top of the atmosphere (TOA). In this study, we show from first principles that temperature feedback within an atmosphere–surface column operates as a set of feedback circuits, each governed by an intrinsic gain less than unity, leading to geometric amplification of energy through successive feedback cycles. In particular, the surface-centered temperature feedback circuit determines the amplification of surface energy per unit TOA forcing. The surface energy amplification by non-temperature feedback is estimated from the amplification of solar energy absorption by downward longwave radiation at the surface in the climate mean state. Climate sensitivity then emerges as the product of these two amplifications scaled by the surface Planck feedback parameter. The climate sensitivity derived from the 1980–2000 climate mean state is 1.15 K (W m$^{-2}$)$^{-1}$. It yields a warming of 0.44 K from 1980–2000 to 2000–2020, consistent with the observed warming of 0.41 K, and equilibrium warmings of 3.02 and 5.83 K for $CO_2$ doubling and quadrupling, respectively, the former consistent with the Charney estimate near 3 °C. The framework provides a basis for predicting global surface temperature responses to other external forcing scenarios from the climate mean state.

**Significance Statement:** Climate sensitivity has been estimated from climate-model simulations, leaving substantial uncertainty in future warming projections. This study demonstrates from first principles that temperature feedback operates as a complete set of feedback circuits within an atmosphere–surface column, leading to geometric amplification of energy perturbations through successive feedback cycles. We prove that temperature-feedback amplification at top-of-the-atmosphere becomes unity at equilibrium, explaining why equilibrium top-of-the-atmosphere feedback analyses cannot reveal such amplification. The climate sensitivity derived from the climate mean state is consistent with the observed global warming and with the Charney estimate of about 3 °C of equilibrium warming for $CO_2$ doubling.

## 1. Introduction

Since the Charney Report (Charney et al. 1979), the equilibrium global warming in response to a doubling of atmospheric $CO_2$ has remained uncertain, with estimates generally ranging from about 2 °C to 5 °C, including several high-sensitivity CMIP6 models with equilibrium climate sensitivity exceeding 5 °C (Collins et al. 2013; Sherwood et al. 2020; IPCC 2021; Forster et al. 2021; Zelinka et al. 2020). This persistent uncertainty is commonly characterized in terms of climate sensitivity, namely the equilibrium change in global mean surface temperature per unit radiative forcing at the top of the atmosphere (TOA). In the literature, the total climate feedback parameter, $\lambda$, is defined as the change in net TOA radiative flux per unit change in global mean surface temperature (Wetherald and Manabe 1988; Soden and Held 2006; Gregory et al. 2004; Bony et al. 2006, 2015). Climate sensitivity can then be expressed as

$$\beta = -\frac{1}{\lambda}. \qquad (1)$$

Within radiative perturbation frameworks, particularly the partial radiative perturbation (PRP) method (Wetherald and Manabe 1988; Soden and Held 2006; Bony et al. 2006), the total climate feedback parameter is further decomposed into contributions from the Planck response and individual feedback processes, such as water vapor (WV), clouds, surface albedo, and lapse-rate feedbacks, which can be written symbolically as

$$\lambda = -\lambda_{Planck} + \lambda_{WV} + \lambda_{cloud} + \lambda_{albedo} + \lambda_{lapse-rate}. \qquad (2)$$

Under the PRP method, external forcing at the TOA is treated as the initial energy input and surface warming as the response required to restore radiative balance through enhanced outgoing longwave radiation (OLR), while non-temperature and lapse-rate feedbacks contribute additional energy perturbations at the TOA. Because both surface warming and the associated TOA radiative perturbations from feedback processes are determined together through the climate response, the individual climate feedback parameters, except for the Planck feedback parameter, are generally diagnosed from climate model simulations (Zelinka et al. 2020; Wetherald and Manabe 1988; Soden and Held 2006; Meehl et al. 2020), observational trend analyses (Forster and Gregory 2006), or regression-based analyses of transient responses (Gregory et al. 2004). As a result, these feedback parameters, as well as the total climate feedback parameter, are fundamentally diagnostic rather than predictive in a first-principles sense.

The main objective of this study is to develop a first-principles quantification of the climate feedback parameter and climate sensitivity from the climate mean state that provides predictive capability without relying on the realized climate response. Specifically, we show that temperature feedback within an atmosphere–surface column operates as a set of feedback circuits, each governed by an intrinsic gain less than unity, leading to geometric amplification of energy through successive feedback cycles. We prove that the feedback gain associated with the surface-centered temperature feedback circuit determines the amplification of surface energy per unit TOA forcing. The amplification of surface energy by non-temperature feedback is estimated from the amplification of solar energy absorption by downward longwave radiation at the surface in the climate mean state. Rather than the additive form shown in Eq. 2, the climate feedback parameter is shown to emerge as

$$\lambda = -\frac{\lambda_{Planck}}{A_{T_s}^{(TOA)} A_{NT}}, \qquad (3)$$

where $\lambda_{Planck} = 4\sigma\bar{T}_s^3$ represents the surface Planck feedback parameter, with $\sigma$ denoting the Stefan-Boltzmann constant and $\bar{T}_s$ the mean surface temperature, $A_{T_s}^{(TOA)}$ denotes the amplification of surface energy by temperature feedback from a unit TOA forcing reaching the surface, and $A_{NT}$ denotes the amplification of surface energy by non-temperature feedback. It is demonstrated that the theoretically derived climate feedback parameter and associated climate sensitivity are consistent with the observed warming, and that the predicted warming for $CO_2$ doubling is remarkably consistent with the original Charney estimate near 3 °C.

## 2. Data and methods

### *a. Data*

The monthly mean data from the European Centre for Medium-Range Weather Forecasts ERA5 reanalysis (Hersbach et al. 2020) covering the period 1980–2000 are used to construct the global-mean climate state, including vertical profiles of atmospheric temperature, specific humidity, clouds, and ozone, together with surface albedo, surface pressure, surface temperature, downward and upward shortwave fluxes at the surface, and downward longwave flux at the surface. The monthly mean ERA5 surface temperature data for 2000–2020 are used to obtain the global-mean surface temperature for 2000–2020. The difference between the 2000–2020 and 1980–2000 global-mean surface temperatures is regarded as the observed global warming and is used to validate the predicted warming. The ERA5 reanalysis is archived at the Climate Data Store. The dataset has 37 pressure levels plus the surface level. The global-mean surface pressure of ERA5 reanalysis is 985.6 hPa. As a result, the global mean atmosphere–surface column used in the calculations consists of 36 atmospheric layers above the surface and one surface layer, giving a total of 37 layers.

*b. Radiative-transfer calculations and the Planck feedback matrix*

We apply the radiative transfer model described in Fu and Liou (1992, 1993) to calculate all radiative quantities used in this study from the prescribed climate mean states. The radiative transfer model outputs, all in units of W m$^{-2}$, include the vertical profiles of shortwave (SW) heating rates, $\left(S_i^{(base)}\right)$, and longwave (LW) cooling rates, $\left(R_i^{(base)}\right)$. Using the same radiative transfer model, we obtain the vertical profiles of SW and LW radiative heating rates at a prescribed $CO_2$ concentration different from that in the baseline climate state, denoted as $\left(S_i^{(CO_2)}\right)$ and $\left(R_i^{(CO_2)}\right)$, respectively. The vertical profile of net radiative heating perturbations due to the $CO_2$ change, denoted by $\Delta\mathbf{F}^{(CO_2)}$, is then obtained as

$$\Delta\mathbf{F}^{(CO_2)} = \left(S_i^{(CO_2)}\right) - \left(R_i^{(CO_2)}\right) - \left[\left(S_i^{(base)}\right) - \left(R_i^{(base)}\right)\right]. \qquad (4)$$

Following Lu and Cai (2009) and Cai and Lu (2009), the Planck feedback matrix, $\left(\partial R_i / \partial T_j\right)$, of size $(L+1) \times (L+1)$ and in units of W m$^{-2}$ K$^{-1}$, is obtained by repeating the radiative transfer model calculations $L+1$ times, where $L$ denotes the total number of atmospheric layers in the atmosphere–surface column, with the first layer representing the top layer of the atmosphere and the surface layer treated as the $(L+1)$th layer. In each calculation, a 1 K temperature perturbation is imposed at the $j$-th layer while all other variables are fixed at their baseline climate mean-state values. The resulting vertical profile of LW cooling rates is denoted by $\left(R_i^{(j)}\right)$. The difference between $\left(R_i^{(j)}\right)$ and the baseline LW cooling profile, $\left(R_i^{(base)}\right)$, gives the vertical profile of radiative cooling perturbations due to the 1 K warming imposed at the $j$-th layer, corresponding to the $j$-th column of the Planck feedback matrix,

$$\frac{\partial R_i}{\partial T_j} = R_i^{(j)} - R_i^{(base)}, \quad i = 1, 2, \ldots, L+1. \qquad (5)$$

Repeating this procedure for $j = 1,2, \ldots, L+1$ yields the full Planck feedback matrix.

**3. Formulation of temperature feedback circuits**

The Planck feedback matrix provides a compact representation of how temperature changes in one layer influence radiative heating throughout the atmosphere–surface column. As illustrated in Fig. 1a, the diagonal elements of $\left(\partial R_i / \partial T_j\right)$ are all positive, indicating longwave (LW) emission perturbations from individual layers due to their 1 K warming, whereas off-diagonal elements, $\partial R_i / \partial T_j < 0$ for $i \neq j$, indicate heating perturbations in other layers due to their LW absorption of emission from the diagonal layer of the corresponding column. The column coupling describes how thermal emission from a given layer propagates vertically. When a layer warms, it emits more radiation, which is partially absorbed by layers above and below. The vertical summation of all elements in a column of the Planck feedback matrix, which is always positive, corresponds to the outgoing longwave radiation (OLR) perturbation at the top of the atmosphere (TOA) resulting from a 1 K warming of the corresponding layer. The off-diagonal elements in a row of the Planck feedback matrix describe the absorption of LW radiation emitted from all other layers by the corresponding layer. The vertical summations of all individual columns of the Planck feedback matrix constitute the temperature kernel (Soden and Held 2006). By construction, the temperature kernel retains only the net emission effect at the TOA associated with temperature feedback, while the absorption effect of LW radiation within the atmosphere–surface column is lost by the vertical summation.

Utilizing the full radiative coupling information contained in the Planck feedback matrix, Cai et al. (2024) derived the energy gain kernel (EGK). The diagonal elements of EGK represent the amplification of energy perturbations at a layer where the unit energy perturbation is imposed, while the off-diagonal elements correspond to energy gained at other layers, both of which arise from radiative coupling within the atmosphere–surface column. The derivation of the EGK is based on the equilibrium solution in response to a unit energy perturbation. As such, it does not explicitly describe the processes through which energy amplification develops, nor how temperature feedback operates to amplify input energy perturbations while the system simultaneously approaches radiative equilibrium. To reveal both the amplification pathway and the attainment of radiative balance, we decompose the temperature feedback within the atmosphere–surface column into a complete set of (L + 1) feedback circuits, one centered on each layer, through which energy perturbations are amplified over successive feedback cycles and converge to the EGK at equilibrium.

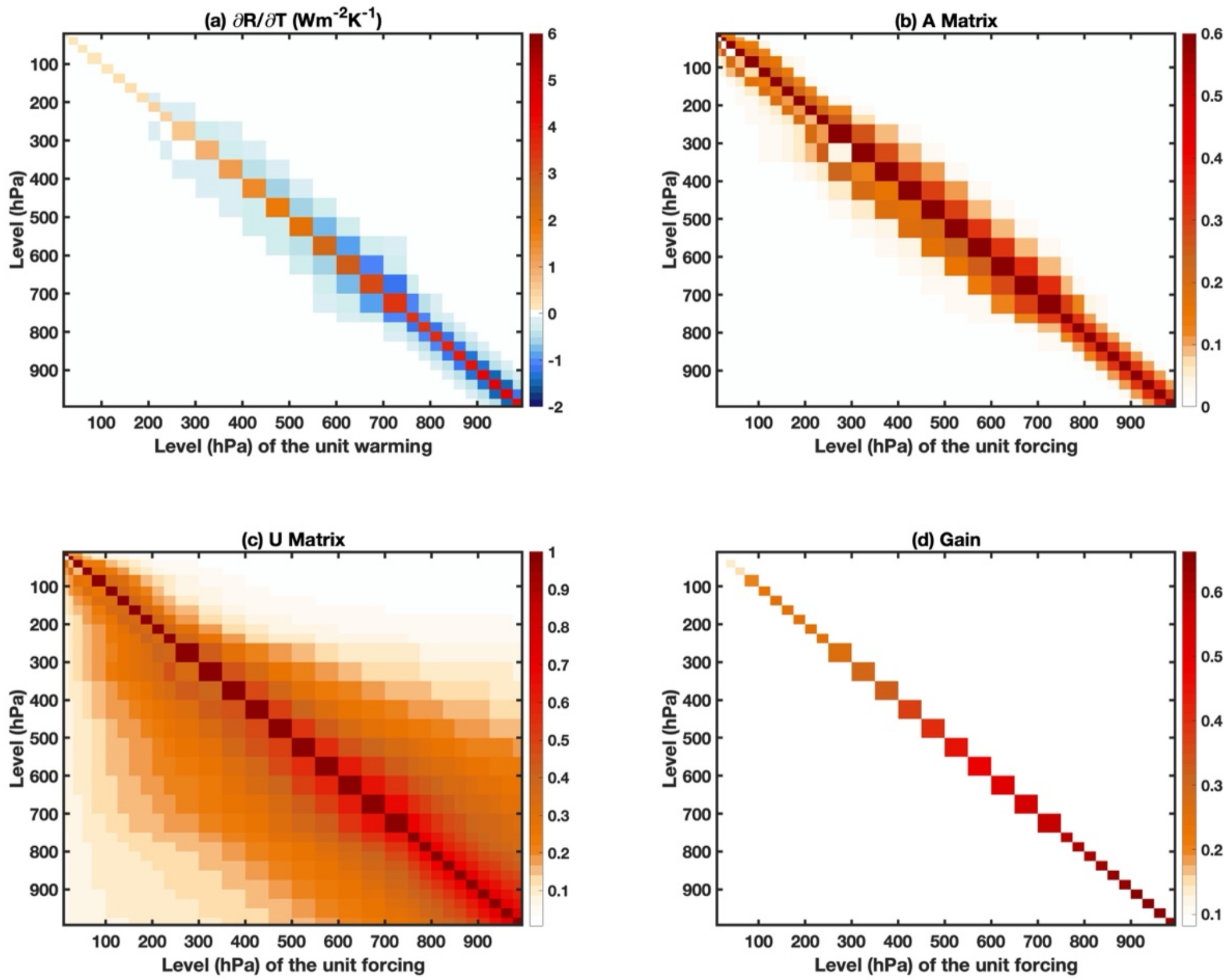

**Fig. 1.** Radiative coupling matrices for temperature feedback circuits. (a) Planck feedback matrix, (b) matrix $\mathbf{A}$, (c) matrix $\mathbf{U}$, and (d) diagonal matrix of intrinsic circuit gains, $\boldsymbol{G} = \mathrm{diag}(g_i)$. The diagonal elements of both $\mathbf{A}$ and $\mathbf{U}$ are equal to unity, while the off-diagonal elements are positive and less than unity. The diagonal elements of $\boldsymbol{G}$ are positive and less than unity, while all off-diagonal elements are zero. The matrices are computed from the ERA5 1980–2000 global-mean atmosphere–surface column state consisting of 37 layers.

Without loss of generality, we consider how an input energy perturbation at layer $i$, referred to as the source layer, is amplified in a compounding manner through successive temperature feedback cycles associated with radiative coupling within the atmosphere–surface column, while the system approaches radiative equilibrium. As illustrated in Fig. 2, the temperature feedback circuits consist of two types of closed circuits. The first is a source-layer-centered circuit, in which the source layer receives an external energy input. This circuit describes how energy emitted from the source layer undergoes radiative coupling with non-source layers, producing successive compounding amplification as it propagates through the atmosphere–surface column and contributes to a compounding increase in outgoing longwave radiation that balances the imposed energy perturbation. The second consists of $L$ non-source-layer-centered circuits, which operate in a manner similar to the source-layer-centered circuit, with each non-source layer $k$ having its own intrinsic circuit gain, except that their initial energy source is internal to the atmosphere–surface column. This energy source arises from the initial energy gain at non-source layers that is not accounted for in the source-layer-centered circuit. Because only the source-layer-centered circuit involves an external energy input, it is the only circuit that contributes directly to the energy balance at the top of the atmosphere. Consequently, radiative equilibrium at the TOA, which balances the imposed energy perturbation, is achieved through the source-layer-centered circuit, while the non-source-layer-centered circuits are driven by internal energy gain within the system and do not directly contribute to the TOA energy balance.

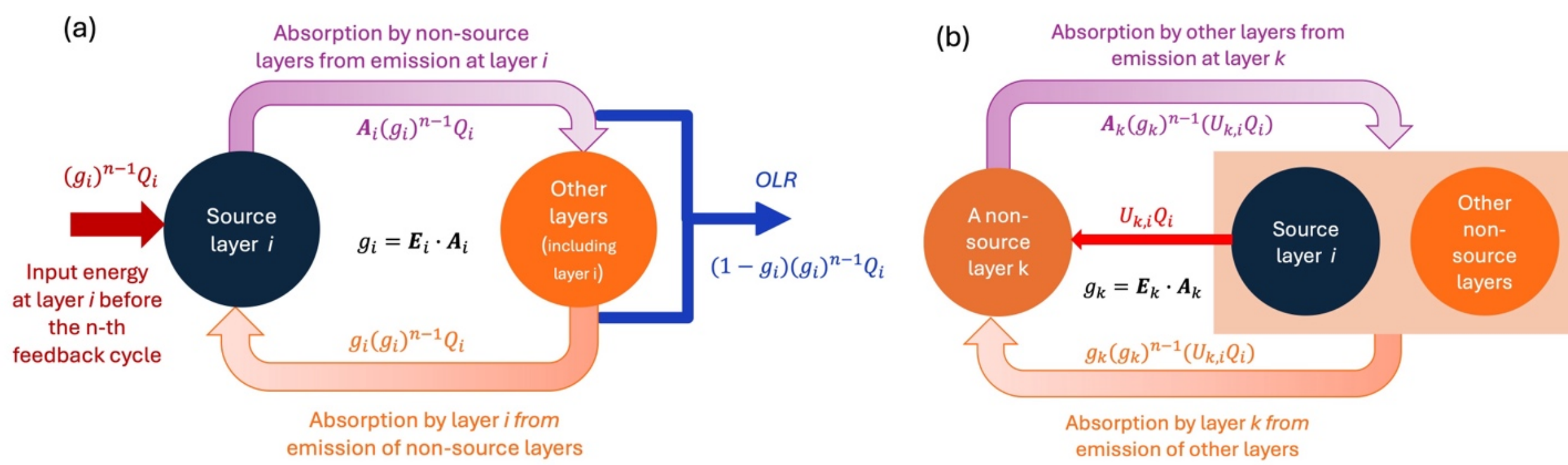

**Fig. 2.** Schematic illustration of the temperature feedback circuits. (a) Source-layer-centered circuit and (b) non-source-layer-centered circuit. The blue circles represent the source layer $i$, where the input energy perturbation $Q_i$ is imposed, while the orange circles represent non-source layers. For the source-layer-centered circuit, the red arrow denotes the unbalanced energy entering layer $i$ prior to the $n$-th feedback cycle, $(g_i)^{n-1}Q_i$, where $g_i = \mathbf{E}_i \cdot \mathbf{A}_i = \sum_{j \neq i} A_{j,i}\, U_{i,j}$, corresponding to the intrinsic circuit gain of layer $i$. The upper pathway represents LW radiation emitted from the source layer $i$ and absorbed by the non-source layers. During the first feedback cycle, the energy absorbed by the non-source layers within this circuit is $\mathbf{A}_i Q_i$, where $\mathbf{A}_i$ denotes the vertical profile of the fraction of energy absorbed by those layers. The lower pathway represents LW radiation emitted from the warmed non-source layers and absorbed by layer $i$. The blue arrow denotes LW radiation emitted to space during the $n$-th cycle. Non-source-layer-centered circuits operate in a similar manner, with each non-source layer $k$ having its own intrinsic circuit gain $g_k = \mathbf{E}_k \cdot \mathbf{A}_k = \sum_{j \neq k} A_{j,k}\, U_{k,j}$, but with an initial energy perturbation, $U_{k,i}Q_i$, that is gained internally from the source layer.

*a. Foundational radiative-coupling analysis for temperature feedback circuits*

The energy gains of the source layer and each non-source layer are determined by two radiative coupling matrices (**A** and **U**) of size $(L+1)\times(L+1)$ for energy gain in the atmosphere–surface column (Figs. 1b and 1c), derived from the Planck feedback matrix. The elements of matrix **A** are given by

$$0 < A_{k,i} = -\frac{\partial R_k / \partial T_i}{\partial R_i / \partial T_i} < 1, \quad k \neq i, \qquad A_{i,i} = 1. \qquad (6)$$

The off-diagonal elements of the $i$th column of matrix **A** correspond to the vertical profile of the fraction of a unit LW emission from layer $i$ absorbed by each other layer $k$, referred to hereafter as the $\mathbf{A}_i$-vertical profile.

To derive matrix **U**, we first consider a reduced system in which layer $i$ is excluded. The Planck feedback matrix for the remaining $L$ layers is otherwise identical to the full Planck feedback matrix except the $i$th row and $i$th column are removed. The temperature response of this reduced system to a unit energy input at each layer $l$ is obtained from the inverse of the reduced Planck feedback matrix, denoted as $\left(\delta T_{k',l'}^{(i)}\right)$, where the superscript $(i)$ denotes that the inverse is obtained without the $i$th row and $i$th column of the full Planck feedback matrix. For notational convenience, we define a mapping that embeds $\left(\delta T_{k',l'}^{(i)}\right)$, which is an $L\times L$ matrix, into an $(L+1)\times(L+1)$ matrix as follows:

$$\widehat{\delta T}_{k,l}^{(i)} = \delta T_{k',l'}^{(i)}, \quad k = \begin{cases} k', & k' < i, \\ k'+1, & k' \geq i, \end{cases} \quad l = \begin{cases} l', & l' < i, \\ l'+1, & l' \geq i. \end{cases} \qquad (7)$$

By construction, the $i$th row and $i$th column of matrix $\left(\widehat{\delta T}_{k,l}^{(i)}\right)$ are undefined. The $i$th row of matrix **U** can be obtained from the product of the $i$th row of the full Planck feedback matrix and $\left(\widehat{\delta T}_{k,l}^{(i)}\right)$,

$$U_{i,l} = -\sum_{k\neq i} \frac{\partial R_i}{\partial T_k} \widehat{\delta T}_{k,l}^{(i)}, \quad l \neq i, \qquad U_{i,i} = 1. \qquad (8)$$

Repeating the above calculation for all $(L+1)$ layers yields the full matrix **U**. As shown in Fig. 1c, all off-diagonal elements of **U** satisfy $0 < U_{i,l} < 1$. The off-diagonal elements of the $i$th row of **U** correspond to the vertical profile of the fraction of LW emission from each non-$i$ layer $l$, after radiative equilibrium is established among all non-$i$ layers, absorbed by layer $i$. For reference, the $i$th row of $\mathbf{U}$ is referred to hereafter as the $\mathbf{E}_i$-vertical profile. The off-diagonal elements of the $i$th column of $\mathbf{U}$ correspond to the vertical profile of the fraction of LW emission from layer $i$ absorbed by each non-$i$ layer $k$, after radiative equilibrium is established among all non-$k$ layers. For reference, the $i$th column of $\mathbf{U}$ is referred to hereafter as the $\mathbf{U}_i$-vertical profile.

*b. Source-layer-centered circuits*

The source-layer-centered circuit (Fig. 2a) begins with emission from the source layer $i$ that balances the imposed input energy. A fraction of this emission is absorbed by other layers, as given by the $\mathbf{A}_i$-vertical profile. The product of the $\mathbf{E}_i$-vertical and $\mathbf{A}_i$-vertical profiles corresponds to the fraction of LW emission perturbation from the non-source layers that is absorbed by layer $i$. This fraction constitutes the additional unbalanced energy entering layer $i$ during each feedback cycle and thus represents the gain of the $i$th-layer-centered circuit, denoted as $g_i = \mathbf{E}_i\mathbf{A}_i$. The remaining fraction of energy that does not enter the source layer $i$ during a feedback cycle is $(1 - g_i)$, which escapes from the atmosphere–surface column as OLR. We note that the condition $g_i = \mathbf{E}_i\mathbf{A}_i < 1$ ensures at least part of the input energy $Q_i$ can be output as OLR through the thermal emission in each feedback cycle. This condition also holds even in the limiting case in which the vertical sum of $\mathbf{A}_i$ equals 1, in which the emission from layer $i$ is completely absorbed by the other layers. Otherwise, $g_i = \mathbf{E}_i\mathbf{A}_i > 1$ would indicate an instability of the radiative coupling within the atmosphere–surface column.

The unbalanced energy returning to layer $i$ after each feedback cycle from the first through the $N$-th feedback cycle follows a geometric sequence

$$g_i Q_i, \quad g_i^2 Q_i, \quad \dots, \quad g_i^N Q_i, \qquad (9)$$

where $g_i$ denotes the intrinsic circuit gain associated with layer $i$, and is given by

$$0 < g_i = \sum_{j \neq i} A_{j,i}\, U_{i,j} < 1, \qquad (10)$$

with matrices $\mathbf{A}$ and $\mathbf{U}$ given in Eqs. (6) and (8), respectively. At the end of each feedback cycle, the unbalanced energy entering the source layer $i$ in the previous cycles is balanced through its emission perturbations. At the same time, the OLR emitted from the atmosphere–surface column during the successive feedback cycles also follows a geometric sequence given by

$$(1-g_i)Q_i, \quad (1-g_i)g_iQ_i, \quad \ldots, \quad (1-g_i)g_i^{N-1}Q_i. \qquad (11)$$

Because $g_i < 1$, the geometric-series representation of the successive feedback cycles converges. Summing over feedback cycles from $n = 1$ to $N$, which represents the cumulative effect of the first N feedback cycles of the circuit, yields the total energy at the source layer $i$ through cycle $N$, $E_i^{(N)}$ (including the initial input energy $Q_i$), and the total OLR perturbation through cycle $N$, $OLR_i^{(N)}$. Specifically,

$$E_i^{(N)} = \sum_{n=1}^{N} g_i^{n-1}\, Q_i = \frac{1-g_i^N}{1-g_i}\, Q_i, \qquad (12)$$

$$\mathrm{OLR}_i^{(N)} = \sum_{n=1}^{N} (1-g_i)\, g_i^{n-1} Q_i = (1-g_i^N) Q_i. \qquad (13)$$

These expressions show that $E_i^{(N)}$ and $\mathrm{OLR}_i^{(N)}$ both increase monotonically with $N$. The increment from one cycle to the next decreases progressively, as it is proportional to $g_i^N$, indicating that the rate of increase diminishes with increasing $N$ and approaches zero as $N \to \infty$.

At equilibrium, as $N \to \infty$, the total energy entering the source layer $i$ that is balanced through its emission perturbations is

$$\frac{\partial R_i}{\partial T_i} \Delta T_i = E_i = \frac{1}{1-g_i} Q_i, \qquad (14)$$

where $\partial R_i / \partial T_i$ is the $i$-th diagonal element of the Planck feedback matrix shown in Fig. 1a, and $\Delta T_i$ denotes the warming at source layer $i$ in response to the amplified energy perturbation $E_i$. The total OLR emitted from the atmosphere–surface column at equilibrium is exactly equal to the imposed input energy $Q_i$, namely,

$$\mathrm{OLR} = \sum_{n=1}^{\infty} (1-g_i)\, g_i^{n-1} Q_i = Q_i. \qquad (15)$$

*c. Non-source-layer-centered circuits*

Non-source-layer-centered circuits operate in a manner similar to the source-layer-centered circuit, with each non-source layer $k$ having its own intrinsic circuit gain determined by the product of its $\mathbf{E}_k$-vertical and $\mathbf{A}_k$-vertical profiles, namely $g_k = \mathbf{E}_k\mathbf{A}_k$. By definition, the initial energy input to non-source layers originates internally from the source layer where the energy input $Q_i$ is imposed. As a result, the energy balance in non-source-layer-centered circuits is achieved internally through radiative coupling within the atmosphere–surface column, without involving OLR, as illustrated in Fig. 2b. The $\mathbf{A}_i$-vertical profile, however, only determines the fraction of energy entering non-source layers from the source layer $i$ under the condition of local equilibrium of the source layer $i$ with its energy input, without fully accounting for the two-way radiative coupling between a given non-source layer and the rest of the atmosphere–surface column. In contrast, $U_{k,i}$ represents the fraction of the energy imposed at the source layer that enters the non-source layer $k$, which then seeds subsequent energy gain through the $k$th-layer-centered circuit with the corresponding intrinsic circuit gain. As indicated in Figs. 1b and 1c, the off-diagonal elements of matrix $\mathbf{U}$ are all greater than their counterparts in matrix $\mathbf{A}$. This is expected because accounting for radiative equilibrium among the source layer and all other non-source layers in response to the imposed energy input allows additional energy gain through radiative coupling, resulting in a larger fraction of energy entering a given non-source layer than that obtained when only the local equilibrium of the source layer is considered.

The unbalanced energy initially entering a non-source layer $k$ is $U_{k,i}Q_i$, where $U_{k,i}$ is given by Eq. (8). Accordingly, the unbalanced energy returning to layer $k$ after each feedback cycle follows a geometric sequence

$$g_k U_{k,i} Q_i, \quad g_k^2 U_{k,i} Q_i, \quad \dots, \quad g_k^N U_{k,i} Q_i, \quad k \neq i. \qquad (16)$$

The total energy at layer $k$ after the $N$ feedback cycles through its feedback circuit is given by

$$E_k^{(N)} = \sum_{n=1}^{N} g_k^{n-1}\, U_{k,i} Q_i = \frac{1 - g_k^N}{1 - g_k} U_{k,i} Q_i, \quad k \neq i. \qquad (17)$$

At equilibrium, as $N \to \infty$, the total energy entering layer $k$ that is balanced through its emission perturbations is

$$\frac{\partial R_k}{\partial T_k} \Delta T_k = E_k = \frac{U_{k,i}}{1 - g_k} Q_i, \quad k \neq i. \qquad (18)$$

*d. Perturbation energy balance in finite-cycle and equilibrium states*

Based on Eqs. (9) and (16), the perturbation energy equation after $N$ feedback cycles for the atmosphere–surface column can be written in matrix form,

$$\mathrm{diag}\left(\frac{\partial R_i}{\partial T_i}\right)\left(\Delta T_i^{(N)}\right) = (\mathbf{I} - \mathbf{G}^N)(\mathbf{I} - \mathbf{G})^{-1}\mathbf{U}\mathbf{Q}, \qquad (19)$$

where

$$\mathbf{G} = \mathrm{diag}(g_1, g_2, \ldots, g_{L+1}). \qquad (20)$$

In Eqs. (19)–(20), $\mathbf{G}$ is the temperature-feedback circuit gain matrix (Fig. 1d), $\mathbf{I}$ is the identity matrix, $\mathbf{Q}$ denotes the vertical profile of energy perturbations induced either externally or internally by non-temperature feedback (including changes in water vapor, clouds, and surface albedo, as well as non-radiative processes such as surface turbulent heat fluxes, convection, and large-scale circulations), and $\left(\Delta T_i^{(N)}\right)$ corresponds to the vertical profile of temperature response to $\mathbf{Q}$ after $N$ feedback cycles. The OLR perturbation emitted through the source-layer-centered circuits up to cycle N is given by

$$\mathrm{OLR}^{(N)} = \mathbf{1}^T(\mathbf{I} - \mathbf{G}^N)\mathbf{Q}, \qquad (21)$$

where $\mathbf{1}^T$ denotes the row vector of ones. At equilibrium ($N \to \infty$), we have

$$\mathrm{diag}\left(\frac{\partial R_i}{\partial T_i}\right)(\Delta T_i) = (\mathbf{I} - \mathbf{G})^{-1}\mathbf{U}\mathbf{Q} = \mathbf{\Gamma}\mathbf{Q}, \qquad (22)$$

where $\mathbf{\Gamma}$ corresponds to the energy gain kernel (EGK), first derived by Cai et al. (2024). And at equilibrium, we have

$$\mathrm{OLR} = \mathbf{1}^T\mathbf{Q} = \sum_i Q_i\,. \qquad (23)$$

The comparison of Eqs. (19) and (22) indicates that, at individual layers, the effect of temperature feedback on imposed energy perturbations follows a multiplication rule in both the finite-cycle and equilibrium states. However, at the TOA, the multiplication rule associated with temperature feedback applies only in the finite-cycle state, as indicated in Eq. (21). In equilibrium, the multiplication factor associated with temperature feedback at the TOA becomes exactly equal to unity, as indicated in Eq. (23). Therefore, equilibrium TOA-based feedback analysis cannot explicitly characterize the internal amplification mechanism associated with temperature feedback, as the amplification information is no longer retained at the TOA once radiative equilibrium is achieved.

## 4. Mapping of TOA forcing to surface energy and its amplification via temperature feedback

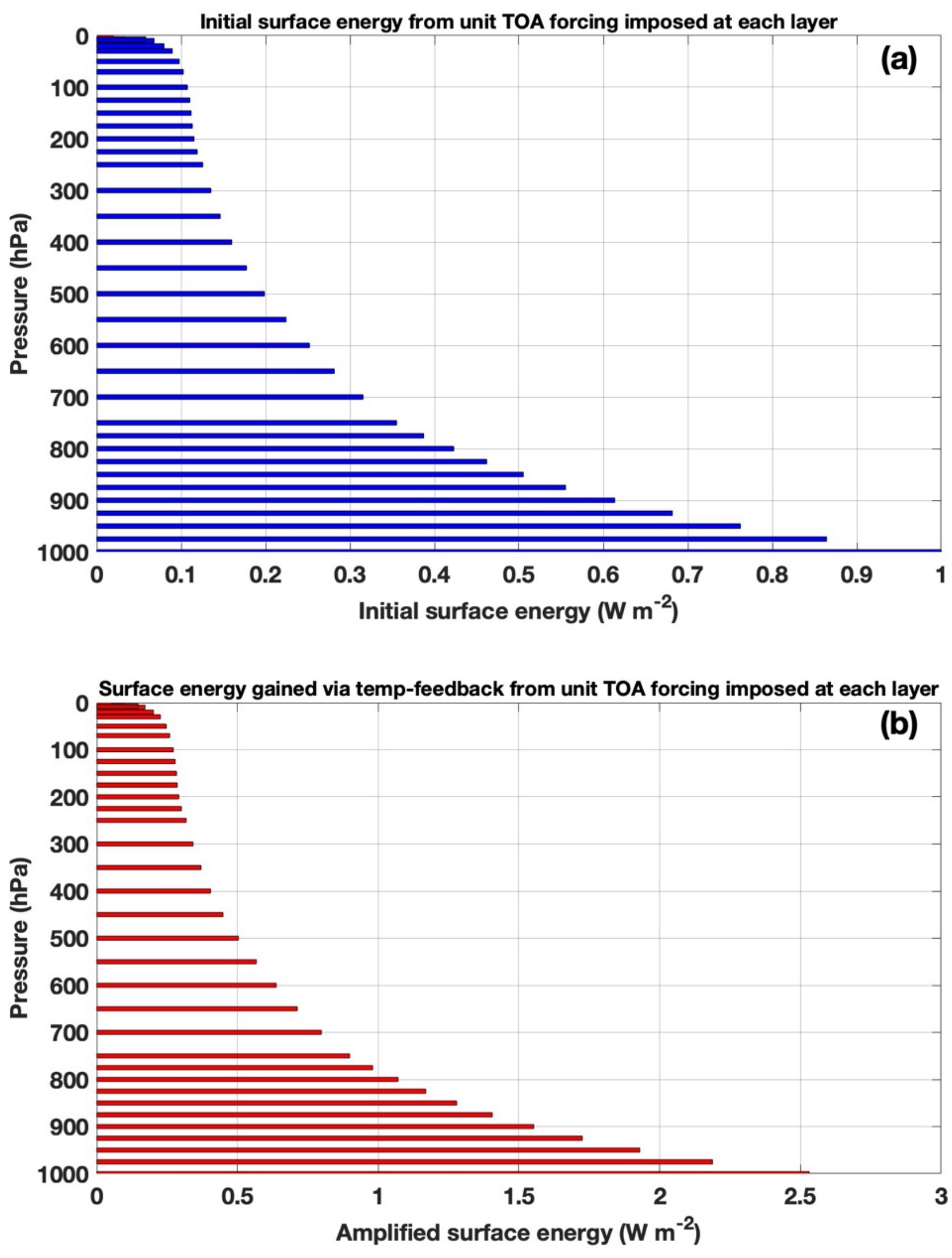

**Fig. 3.** Vertical profiles of surface energy associated with a unit TOA forcing imposed at individual layers of the global–mean atmosphere–surface column. (a) Fraction of the energy imposed at each layer that enters the surface layer prior to amplification by the surface-centered temperature feedback circuit, given by the surface row of matrix $\mathbf{U}$, $U_{L+1,l}$ (Fig. 1c). (b) Corresponding surface energy after amplification by the surface-centered temperature feedback circuit from the same unit TOA forcing applied at each layer.

We next apply a unit $(1 \text{ W m}^{-2})$ energy perturbation to each of the $(L+1)$ layers, which corresponds to a total TOA forcing of $(L+1)$ W m$^{-2}$. The response to the unit energy perturbation at each layer describes how energy imposed at that layer reaches the surface; by linearity, the result for any forcing profile follows by superposition (section 7). Figure 3a shows the fraction of this energy entering the surface layer from each of the $(L+1)$ layers, obtained from the $(L+1)$th (surface) row of matrix $\mathbf{U}$ shown in Fig. 1c. Specifically, $0 < U_{L+1,l} < 1$ for all $1 \leq l \leq L$ represents the fraction of unbalanced energy gained by the surface layer from an atmospheric layer $l < L+1$, with the surface acting as a non-source layer, whereas $U_{L+1,L+1} = 1$ corresponds to the unit unbalanced energy perturbation imposed directly at the surface, where the surface serves as the source layer. Once unbalanced energy enters the surface layer, it undergoes the same amplification through the surface-centered feedback circuit. In other words, this amplification applies to all $U_{L+1,l}$, regardless of whether the surface acts as a source layer or as a non-source layer. Consequently, each component of the surface energy shown in Fig. 3a is amplified through the same surface-layer circuit gain $g_{L+1}$, leading to the amplified surface energy shown in Fig. 3b.

Therefore, the temperature feedback amplification of surface energy from a unit TOA forcing that reaches the surface, either directly (forcing imposed at the surface) or indirectly (energy transferred from the forcing imposed on the atmosphere), is

$$A_{T_s}^{(TOA)} = \frac{\sum_{l=1}^{L+1} U_{L+1,l} /(1 - g_{L+1})}{\sum_{l=1}^{L+1} U_{L+1,l}} = \frac{1}{1 - g_{L+1}} = \frac{1}{1 - g_{surface}}. \qquad (24)$$

According to Fig. 3a, of the total TOA forcing of 37 W m$^{-2}$ imposed at the 37 layers, 1 (W m$^{-2}$) $\times \sum_{l=1}^{L+1} U_{L+1,l} = 9.94$ W m$^{-2}$ is initially mapped to the surface. From Fig. 3b, this energy is amplified to 1 (W m$^{-2}$) $\times \sum_{l=1}^{L+1} U_{L+1,l} /(1 - g_{surface}) = 25.16 \text{ W m}^{-2}$ through the surface-centered temperature feedback circuit, which leads to the temperature-feedback amplification factor of 25.16/9.94 = 2.53, identical to $\frac{1}{1-g_{surface}}$ given by Eq. (24) with $g_{surface} = 0.605$. This amplification of 2.53 applies to the portion of the TOA forcing that reaches the surface (we note that the exact fraction of a given TOA forcing that reaches the surface depends on its vertical profile, as discussed in section 7).

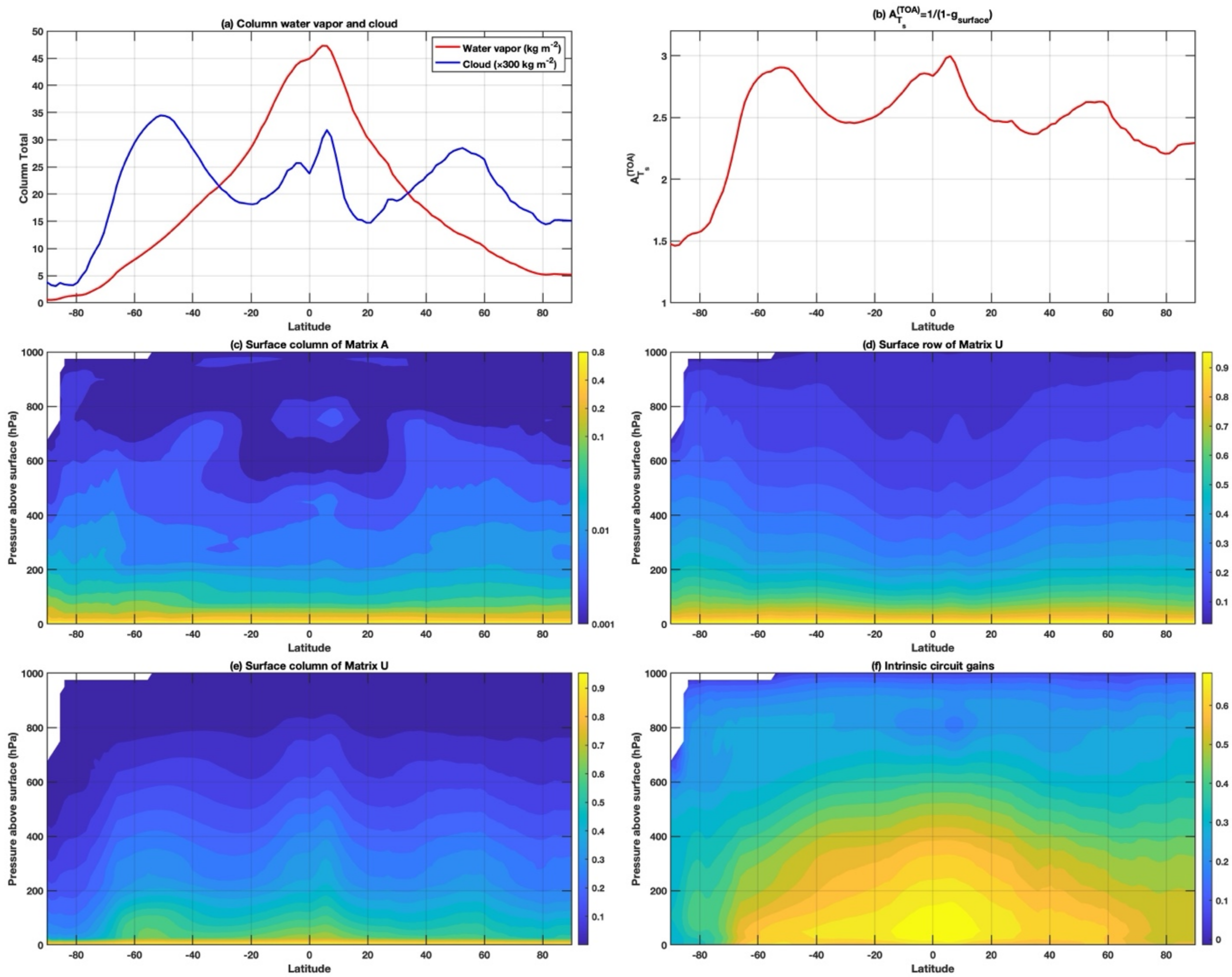

**Fig. 4.** Spatial patterns of longwave absorbers and the strength of temperature feedback circuits. (a) Latitudinal profiles of column-integrated water vapor (red; kg m$^{-2}$) and clouds (blue; scaled by a factor of 300 for visualization). Latitudinal variations of (b) $\mathrm{A}_{\mathrm{T_S}}^{(\mathrm{TOA})} = 1/(1 - g_{surface})$, corresponding to the temperature-feedback amplification of the fraction of the TOA forcing that reaches the surface, (c) surface column of matrix **A**, (d) surface row of matrix **U**, (e) surface column of matrix **U**, and (f) intrinsic circuit gains of temperature feedback. Note that the color scale in (c) is logarithmic. The results are derived from the ERA5 1980–2000 zonal-mean climatological state. The coordinate of pressure thickness above the ground, given by $(p_s - p)$ in units of hPa, where $p_s$ is the surface pressure and $p$ is the pressure at each level, is used so that zonal averaging is performed at the same level relative to the ground, thereby accurately preserving the vertical structure of longwave radiative coupling within the atmosphere–surface column.

The mapping and amplification of surface energy shown in Fig. 3 and Eq. (24) depend strongly on the climate mean state. Figure 4 shows that temperature-feedback amplification exhibits a strong dependence on climate mean states, particularly on the spatial distributions of atmospheric water vapor and clouds. The latitudinal variation of $\boldsymbol{A}_{\boldsymbol{T_s}}^{(\boldsymbol{TOA})}$ (Fig. 4b) is shaped strongly by the latitudinal variations of column-integrated water vapor and clouds (Fig. 4a). The surface-centered temperature feedback circuit gain is determined by the inner product of the surface column of matrix **A**, or the $\mathbf{A}_{\boldsymbol{surface}}$-vertical profile (Fig. 4c), and the surface row of matrix **U**, or the $\mathbf{E}_{surface}$-vertical profile (Fig. 4d). Matrix $\mathbf{A}$ only characterizes the initial absorption effect of the thermal emission resulting from temperature changes in the source layer, as it is directly obtained from the Planck feedback matrix indicated in Eq. (6). Matrix $\mathbf{U}$, on the other hand, already incorporates the temperature feedback effect due to temperature changes in other layers. As a result, the spatial pattern of matrix $\mathbf{U}$ is more strongly shaped by the spatial pattern of atmospheric water vapor and clouds than matrix $\mathbf{A}$, as evident from the $\mathbf{E}_{surface}$-vertical profile shown in Fig. 4d and the $\mathbf{U}_{surface}$-vertical profile shown in Fig. 4e.

## 5. Approximation of non-temperature-feedback amplification from the climate mean state

We now seek a leading-order approximation of non-temperature-feedback amplification in the perturbed climate state from the surface energy balance in the climate mean state. As shown in Eqs. (19), (22) and (24), the relationship between temperature feedback and energy perturbations follows a multiplicative form. Therefore, the surface energy balance for the perturbed climate can be expressed symbolically as

$$4\sigma\bar{T}_s^3 \Delta T_s = A_{T_s}^{(TOA)} A_{NT}\, \Delta F^{(TOA)}, \qquad (25)$$

with

$$A_{NT}\, \Delta F^{(TOA)} = \Delta F^{(TOA)} + \Delta F_s^{(non-temp)}, \qquad (26)$$

where $\bar{T}_s$ denotes the global-mean surface temperature in the climate mean state; $\Delta F^{(TOA)}$ is the imposed TOA energy perturbation; $\Delta F_s^{(non-temp)}$ denotes the surface energy perturbation due to non-temperature feedback; $A_{NT}$ is the amplification factor of non-temperature feedback; and $\Delta T_s$ is the global-mean surface temperature response. The left-hand side of Eq. (25) is the perturbation of the surface energy output through thermal emission. In Eq. (26), we denote the total energy perturbation entering the surface prior to temperature-feedback amplification as $A_{NT}\Delta F^{(TOA)}$. Therefore, the right-hand side of Eq. (25) corresponds to the total input energy perturbation at the surface after temperature feedback amplification.

To estimate $A_{NT}$, we turn to the global-mean surface energy balance in the climate mean state, which is written as

$$\bar{R}_s^{\uparrow} + \bar{Q}_{LH} + \bar{Q}_{SH} = \bar{S}_s^{\downarrow} - \bar{S}_s^{\uparrow} + \bar{R}_s^{\downarrow}, \qquad (27)$$

where $\bar{R}_s^{\uparrow}$ is the surface thermal radiative emission in the climate mean state; $\bar{Q}_{LH}$ and $\bar{Q}_{SH}$ are the surface latent and sensible heat fluxes, respectively (positive upward); $\bar{S}_s^{\downarrow} - \bar{S}_s^{\uparrow}$ represents the solar energy absorbed at the surface; and $\bar{R}_s^{\downarrow}$ is the downward LW flux at the surface from the atmosphere. The left-hand side of Eq. (27) is the total energy output of the surface, and the right-hand side is the total energy input to the surface in the climate mean state. The term $\bar{S}_s^{\downarrow} - \bar{S}_s^{\uparrow}$ can be regarded as the external energy input in the form of absorbed solar radiation, whereas the term $\bar{R}_s^{\downarrow}$ corresponds to an internal source of energy whose values and spatial patterns are determined by the LW absorbers in the climate mean state, including $CO_2$, water vapor, and clouds, and by the atmospheric temperatures, which are in turn determined by both radiative and non-radiative processes (e.g., atmospheric convection and large-scale circulations).

Equations (25) and (27) can be related by analogy. However, Eq. (27) is additive, whereas we write the surface energy balance for the perturbed climate, Eq. (25), in a multiplicative form. Although the perturbed surface energy balance can equally be written in an additive form, the multiplicative form is chosen because it follows from the perturbation energy balance that explicitly relates the temperature response to the energy perturbations through the temperature feedback circuits (i.e., Eqs. 19 and 22). As indicated in Eq. (24), the surface-centered circuit amplifies all energy entering the surface by the same factor, regardless of whether it originates from the external forcing or from non-temperature feedback. As a result, this choice allows $A_{NT}$ to be estimated from the climate mean state, so that the surface temperature response can be predicted before the perturbed climate is known, rather than diagnosed after the response has occurred, as in an additive feedback decomposition.

Because the two forms represent the absence of a process differently, by a zero term in the additive form and by a unit factor in the multiplicative form, $A_{NT}$ cannot be read directly from Eq. (27). To extract information on non-temperature-feedback amplification of surface energy from the additive energy balance in the climate mean state, following Cai et al. (2025), it is a useful intuitive reasoning exercise to consider a hypothetical scenario in which non-temperature feedback is absent. In this scenario, we postulate that the surface energy balance in the climate mean state would reduce to

$$\frac{\bar{R}_s^{\uparrow} + \bar{Q}_{LH} + \bar{Q}_{SH}}{\bar{S}_s^{\downarrow} - \bar{S}_s^{\uparrow}} = 1.$$

In this scenario, the role of temperature feedback is to adjust surface temperature such that the external energy input is balanced through both thermal emission and turbulent fluxes at the surface.

With this hypothetical scenario as a reference, the actual surface energy balance in the climate mean state, Eq. (27), can be written as

$$\frac{\bar{R}_s^{\uparrow} + \bar{Q}_{LH} + \bar{Q}_{SH}}{\bar{S}_s^{\downarrow} - \bar{S}_s^{\uparrow}} = 1 + \frac{\bar{R}_s^{\downarrow}}{\bar{S}_s^{\downarrow} - \bar{S}_s^{\uparrow}}. \qquad (28)$$

Relative to the hypothetical scenario, the surface in the climate mean state outputs additional energy per unit external energy input, given by the second term on the right-hand side of Eq. (28), $\frac{\bar{R}_s^{\downarrow}}{\bar{S}_s^{\downarrow} - \bar{S}_s^{\uparrow}}$, which is the ratio of the two energy sources of the surface in the mean state. The numerator is the downward LW radiation, an internal source that the atmosphere supplies through downward LW emission by atmospheric absorbers (water vapor, clouds, $CO_2$, and other trace gases) at the mean-state atmospheric temperatures. The denominator is the net SW radiation, the external source. The term therefore measures how much energy the atmosphere supplies to the surface per unit of external energy absorbed at the surface. We note that in the additive form, a positive value for $\frac{\bar{R}_s^{\downarrow}}{\bar{S}_s^{\downarrow} - \bar{S}_s^{\uparrow}}$ corresponds to amplification, whereas in the multiplicative form, a value of $A_{NT}$ exceeding unity corresponds to amplification. As shown in Cai et al. (2025, Figs. 2a, 4a and S2), the diagnosed values of $\frac{\bar{R}_s^{\downarrow}}{\bar{S}_s^{\downarrow} - \bar{S}_s^{\uparrow}}$ are greater than one at nearly all grid points in both observations and CMIP6 models, with global-mean values ranging from 1.73 to 2.62 among CMIP6 models, making this ratio a robust candidate for estimating the amplification of surface energy by non-temperature feedback from the climate mean state. Accordingly, we adopt the same approximation as in Cai et al. (2025) to obtain $A_{NT}$, namely,

$$A_{NT} \approx \frac{\bar{R}_s^{\downarrow}}{\bar{S}_s^{\downarrow} - \bar{S}_s^{\uparrow}}. \qquad (29)$$

The validity of Eq. (29) is further supported by extensive numerical evidence reported in Cai et al. (2025). Specifically, the map-correlation skill of $\frac{\bar{R}_s^{\downarrow}}{\bar{S}_s^{\downarrow} - \bar{S}_s^{\uparrow}}$ diagnosed from the climate mean state against $A_{NT}$ diagnosed from the corresponding perturbed climate simulations ranges from about 0.82 to 0.93 for individual CMIP6 models and is 0.91 for their ensemble mean (Fig. S1 in Cai et al. 2025). For the observed climate, $\frac{\bar{R}_s^{\downarrow}}{\bar{S}_s^{\downarrow}-\bar{S}_s^{\uparrow}}$ derived from the 1980–2000 mean state has a map correlation of 0.76 with $A_{NT}$ diagnosed from the change between 1980–2000 and 2000–2020. Using Eq. (29) to estimate $A_{NT}$, Cai et al. (2025) also reproduced the observed global-mean warming and Arctic warming amplification, as well as those in the CMIP6 abrupt-4×$CO_2$ simulations. These results provide strong empirical support for Eq. (29) as a leading-order approximation to the non-temperature-feedback amplification factor in the perturbed climate state.

## 6. Theoretically derived climate feedback parameter and climate sensitivity

Having derived the amplification of energy perturbations from the TOA forcing by temperature feedback and estimated the surface energy amplification by non-temperature feedback from the ratio of the downward LW radiation to the net solar energy absorbed by the surface in the climate mean state, we can directly evaluate climate sensitivity for a given climate mean state and predict the global-mean surface temperature response to an external forcing at the TOA, $\Delta F^{(TOA)}$. Specifically, we can rewrite Eq. (25) as

$$\Delta T_s = \beta_{TOA}\, \Delta F^{(TOA)}, \qquad (30)$$

where

$$\beta_{TOA} = \frac{A_{T_s}^{(TOA)} A_{NT}}{4\sigma \bar{T}_s^3} \qquad (31)$$

is the climate sensitivity. The expression for the climate feedback parameter $\lambda$ given in Eq. (3) then follows directly by applying the definition given in Eq. (1) to Eq. (31).

Table 1 summarizes the quantities that determine $\beta_{TOA}$ for the ERA5 1980–2000 global-mean atmosphere–surface column. The surface-centered circuit gain, $g_{surface} = 0.605$, is the surface element of the intrinsic circuit gains (Fig. 1d). The corresponding temperature-feedback amplification, $A_{T_s}^{(TOA)} = 1/\left(1 - g_{surface}\right) = 2.53$, equals the ratio of the vertically summed amplified surface energy (Fig. 3b) to the vertically summed energy entering the surface (Fig. 3a), as given by Eq. (24). The non-temperature-feedback amplification, $A_{NT} = 2.44$, is obtained from Eq. (29) as the area-weighted global mean of the ratio $\bar{R}_s^{\downarrow}/\left(\bar{S}_s^{\downarrow} - \bar{S}_s^{\uparrow}\right)$ computed at each grid point.

With the surface Planck feedback parameter $4\sigma\bar{T}_s^3 = 5.37$ W m$^{-2}$ K$^{-1}$ for $\bar{T}_s = 287.23$ K, Eq. (31) gives $\beta_{TOA} = 1.15$ K (W m$^{-2}$)$^{-1}$, and the corresponding climate feedback parameter is $\lambda = -0.87$ (W m$^{-2}$ K$^{-1}$).[1] In other words, each unit of TOA forcing reaching the surface results in about 6.2 W m$^{-2}$ of surface energy after amplification by temperature and non-temperature feedbacks ($2.53 \times 2.44$), which is balanced by the increase in surface thermal emission of 5.37 W m$^{-2}$ per kelvin of surface warming. We emphasize that all quantities in Table 1, and hence $\beta_{TOA}$, are derived solely from the 1980–2000 climate mean state, without any information on the forcing or the climate response.

**Table 1.** Climate sensitivity and climate feedback parameter derived from the ERA5 1980–2000 global-mean state

| Quantity | Expression | Value |
|---|---|---|
| Global-mean surface temperature | $\bar{T}_s$ | 287.23 K |
| Surface Planck feedback parameter | $4\sigma\bar{T}_s^3$ | 5.37 W m$^{-2}$ K$^{-1}$ |
| Surface-centered circuit gain | $g_{surface}$ | 0.605 |
| Temperature-feedback amplification | $A_{T_s}^{(TOA)} = 1/(1 - g_{surface})$ | 2.53 |
| Non-temperature-feedback amplification | $A_{NT} \approx \bar{R}_s^{\downarrow}/(\bar{S}_s^{\downarrow} - \bar{S}_s^{\uparrow})$ | 2.44 |
| Climate sensitivity | $\beta_{TOA} = A_{T_s}^{(TOA)} A_{NT}/(4\sigma\bar{T}_s^3)$ | 1.15 K (W m$^{-2}$)$^{-1}$ |
| Climate feedback parameter | $\lambda = -4\sigma\bar{T}_s^3/(A_{T_s}^{(TOA)} A_{NT})$ | −0.87 W m$^{-2}$ K$^{-1}$ |

## 7. Prediction of global-mean surface warming from $CO_2$ forcing

Equation (30) predicts the global-mean surface warming by applying $\beta_{TOA}$ to the TOA forcing. The $CO_2$ forcing, however, is not imposed uniformly across layers but varies in the vertical. Figure 5 shows the vertical profile of the global-mean $CO_2$ forcing, cumulatively integrated from the surface to a given pressure level, for the $CO_2$ increase from 1980–2000 to 2000–2020, $CO_2$ doubling, and $CO_2$ quadrupling, derived from the ERA5 1980–2000 global-mean atmosphere–surface column with a baseline $CO_2$ concentration of 352.2 ppm. In all three cases, the cumulative forcing increases from the surface to about 600 hPa, indicating that the positive forcing is concentrated in the lower troposphere. It then decreases from about 600 hPa to 300–400 hPa, increases again in the upper troposphere, and decreases from about 100 hPa to the TOA. The decreases from 600 hPa to 300–400 hPa and above the tropopause reflect the negative $CO_2$ forcing in these layers. This two-pair positive–negative vertical structure of the $CO_2$ forcing is a robust feature found at all latitudes (Lu and Cai 2010; Cai and Tung 2012; Taylor et al. 2013; Hu et al. 2025). In particular, the negative $CO_2$ forcing in the stratosphere makes the TOA forcing (magenta dotted lines) substantially smaller than the forcing near the tropopause (red dotted lines), as listed in the first two columns of Table 2.

---

[1] The climate feedback parameter λ here is derived from the climate mean state and represents the change in thermal emission at the TOA per unit surface warming. It differs from the climate feedback parameter commonly reported in the literature, which is diagnosed from simulated or observed responses with the forcing defined as the stratospherically adjusted forcing at the tropopause or as the effective radiative forcing; the two values are therefore not directly comparable (see Fig. 5 and Table 2 for the difference between the TOA and tropopause $CO_2$ forcings).

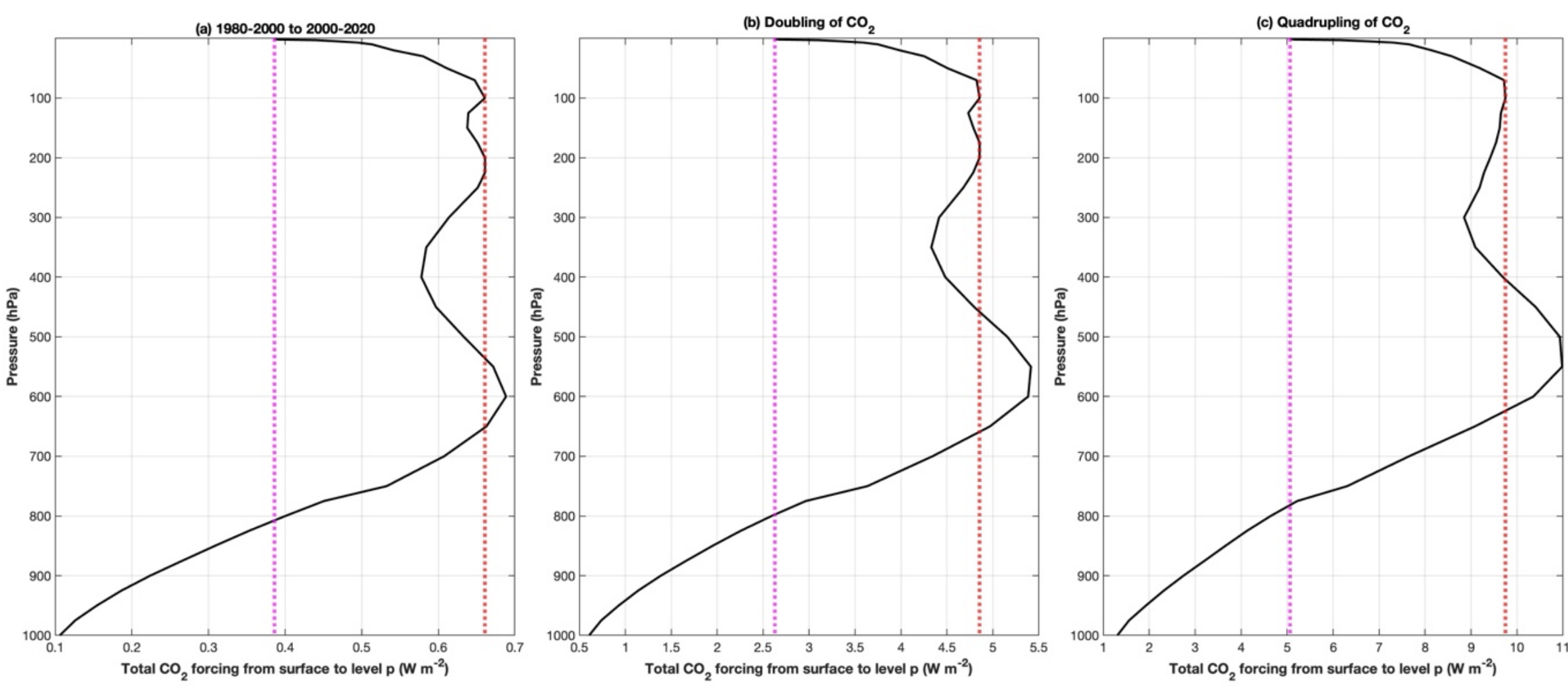

**Fig. 5.** Vertical profiles of global-mean cumulative $CO_2$ radiative forcing from the surface to pressure level $p$. (a) Forcing associated with the increase in $CO_2$ from 352.2 ppm in 1980–2000 to 385 ppm in 2000–2020; (b) forcing for a doubling of $CO_2$ ($2 \times 352.2$ ppm); and (c) forcing for a quadrupling of $CO_2$ ($4 \times 352.2$ ppm). All forcings are calculated from the ERA5 1980–2000 global-mean atmosphere–surface column with a baseline $CO_2$ concentration of 352.2 ppm. The magenta dotted line in each panel indicates the instantaneous $CO_2$ forcing at the TOA, and the red dotted line indicates the instantaneous $CO_2$ forcing at 100 hPa near the tropopause.

The difference between the TOA and tropopause forcings is relevant to the two closely related forms in which external radiative forcing has been defined in the literature. The first is the instantaneous forcing at the TOA, which represents the radiative flux perturbation of the full atmospheric column immediately following a change in $CO_2$, with all climate variables held unchanged (Ramanathan and Dickinson 1979; Hansen et al. 1981; Myhre et al. 1998). The second is the stratospherically adjusted forcing, defined at the tropopause after the stratospheric temperature is allowed to relax to radiative equilibrium while all climate variables in the troposphere and at the surface are held unchanged (Shine et al. 1990; Houghton et al. 2001; Forster et al. 2016). In the stratospherically adjusted forcing, the stratospheric temperature response to the negative stratospheric $CO_2$ forcing is incorporated into the forcing itself. As demonstrated in section 4 and below, the strength of the external energy entering the surface prior to any feedback amplification is represented by the instantaneous TOA forcing, $\Delta F^{(TOA)}$.

**Table 2.** $CO_2$ forcing, climate sensitivity, and predicted global-mean surface warming relative to the 1980–2000 state (352.2 ppm). Forcings are in W m$^{-2}$, climate sensitivities in K (W m$^{-2}$)$^{-1}$, and warming in K. $\Delta F^{(trop)}$ is evaluated at 100 hPa.

| $CO_2$ (ppm) | $\Delta F^{(TOA)}$ | $\Delta F^{(trop)}$ | $\Delta T_s(\beta_{TOA})$ | $\mu$ | $\beta_{profile}$ | $\Delta T_s(\beta_{profile})$ | Observed $\Delta T_s$ |
|---|---|---|---|---|---|---|---|
| 385 | 0.386 | 0.661 | 0.44 | 0.88 | 1.01 | 0.39 | 0.41 |
| 2 × 352.2 | 2.63 | 4.86 | 3.02 | 0.90 | 1.04 | 2.73 | – |
| 4 × 352.2 | 5.07 | 9.75 | 5.83 | 0.92 | 1.05 | 5.33 | – |

According to Eqs. (25) and (30), we apply $\beta_{TOA}$ to the TOA $CO_2$ forcing shown in the second column of Table 2, instead of $\Delta F^{(trop)}$ shown in the third column, to predict the surface temperature response to the $CO_2$ forcing. Applying $\beta_{TOA}$ [1.15 K (W m$^{-2}$)$^{-1}$] derived from the 1980–2000 climate mean state to $\Delta F^{(TOA)} = 0.386$ W m$^{-2}$, resulting from the $CO_2$ increase from 352.2 ppm in 1980–2000 to 385 ppm in 2000–2020, yields a predicted warming of 0.44 K from 1980–2000 to 2000–2020, in close agreement with the observed warming of 0.41 K (Table 2). Equivalently, the observed climate sensitivity, diagnosed as

$$\beta_{obs} = \frac{\bar{T}_s^{2000-2020} - \bar{T}_s^{1980-2000}}{\Delta F^{(TOA)}} = 1.06 \text{ K } \left(\text{W m}^{-2}\right)^{-1}, \qquad (32)$$

is close to $\beta_{TOA}$, which is obtained solely from the 1980–2000 climate mean state. We note that this comparison considers the $CO_2$ forcing only and does not account for other forcings or for the ocean heat uptake associated with the transient climate response; it therefore serves as a consistency check rather than a strict validation. For the idealized perturbations, the predicted equilibrium warming is 3.02 K for $CO_2$ doubling and 5.83 K for $CO_2$ quadrupling, based on TOA forcings of 2.63 and 5.07 W m$^{-2}$, respectively.

Equation (30) applies the temperature-feedback amplification directly to the TOA forcing, without regard to its vertical structure. Because $U_{L+1,l}$ is the fraction of the energy imposed at layer $l$ that enters the surface (section 4), the vertical structure of the forcing can be taken into account through the unbalanced energy entering the surface per unit TOA forcing,

$$\mu = \frac{1}{\Delta F^{(TOA)}} \sum_{l=1}^{L+1} U_{L+1,l}\ \Delta F_l. \qquad (33)$$

Since all energy entering the surface is amplified by the same factor, $A_{T_s}^{(TOA)}$, regardless of the layer from which it enters [Eq. (24)], the climate sensitivity that accounts for the vertical structure of the forcing is

$$\beta_{profile} = \mu\, \beta_{TOA}. \qquad (34)$$

Like $\beta_{TOA}$, $\beta_{profile}$ is obtained without any information on the climate response, because $\mu$ requires only matrix **U** of the climate mean state and the vertical profile of the forcing. We note that $\beta_{profile} \Delta F^{(TOA)}$ is identical to the expression based on the surface row of the EGK used in Cai et al. (2025) to predict both the observed global warming and its spatial pattern, because the surface row of $\mathbf{\Gamma}$ in Eq. (22) is $U_{L+1,j}/(1 - g_{surface})$. When $\beta_{profile}$ is evaluated at each grid point from the local climate mean state and applied to the local forcing profile, it predicts the spatial pattern of global warming, including Arctic amplification, as demonstrated in Cai et al. (2025).

For the three $CO_2$ cases, $\mu$ ranges from 0.88 to 0.92 (Table 2). The value of μ close to unity confirms that the TOA forcing represents the strength of the energy entering the surface. The positive forcing in the lower troposphere is only partially mapped to the surface ($U_{L+1,l} < 1$), whereas the negative forcing in the stratosphere, which reduces the TOA forcing, is mapped to the surface only weakly; these two effects largely offset each other. Because μ is close to unity, $\beta_{TOA}$, which does not take the vertical structure of the forcing into account, provides a good approximation, within about 10%, to $\beta_{profile}$, which ranges from 1.01 to 1.05 K (W m$^{-2}$)$^{-1}$. The warming predicted from $\beta_{profile}$ is 0.39 K for the observed $CO_2$ increase, 2.73 K for $CO_2$ doubling, and 5.33 K for $CO_2$ quadrupling. These results demonstrate that using the total TOA forcing provides a good approximation for predicting global-mean surface warming, with a correction of about 10% when the vertical distribution of $CO_2$ forcing is explicitly accounted for. This correction reflects the vertical mapping of imposed forcing to the surface rather than a change in the intrinsic amplification of the surface-centered temperature-feedback circuit.

The observed climate sensitivity, $\beta_{obs} = 1.06$ K (W m$^{-2}$)$^{-1}$, lies between $\beta_{profile}$ and $\beta_{TOA}$, and the observed warming of 0.41 K lies between the warming predicted from the two. For $CO_2$ doubling, the predicted warmings of 2.73 ($\beta_{profile}$) and 3.02 K ($\beta_{TOA}$) are both consistent with the original assessment of the Charney Report (Charney et al. 1979), which identified "the most probable global warming for a doubling of $CO_2$ to be near 3 °C."

## 8. Concluding remarks

Utilizing the full radiative-coupling information contained in the Planck feedback matrix, this study demonstrates that temperature feedback operates as a complete set of feedback circuits within an atmosphere–surface column, one centered on each layer, each governed by an intrinsic positive circuit gain smaller than unity. Energy perturbations are amplified geometrically through successive feedback cycles, both at the layers where energy is imposed and at other layers to which energy is transferred through radiative coupling, while the system approaches radiative equilibrium. At equilibrium, the resulting amplification converges to the energy gain kernel. At individual layers, temperature feedback amplifies imposed energy perturbations through a multiplication rule in both the finite-cycle and equilibrium states. At the TOA, however, the multiplication factor becomes exactly unity at equilibrium. This explains why traditional equilibrium TOA-based feedback analyses cannot explicitly reveal the amplification associated with temperature feedback, because the amplification information is no longer retained at the TOA once radiative equilibrium is achieved.

The TOA forcing imposed at individual layers reaches the surface either directly, when imposed at the surface, or indirectly, through energy transferred from the atmospheric layers where it is imposed via radiative coupling. We prove that the energy reaching the surface is amplified through the surface-centered temperature feedback circuit by an amount determined solely by the intrinsic circuit gain of the surface layer, regardless of the layer at which the forcing is imposed and regardless of whether the energy originates from external forcing or from non-temperature feedback. The surface intrinsic circuit gain, which increases with the abundance of atmospheric longwave absorbers, particularly water vapor and clouds, therefore determines the temperature-feedback amplification of the surface energy resulting from TOA forcing. The surface energy amplification by non-temperature feedback is estimated from the amplification of solar energy absorption by downward longwave radiation at the surface in the climate mean state. Climate sensitivity then emerges as the product of these two amplifications scaled by the surface Planck feedback parameter, and is therefore determined by the climate mean state alone. Applying this climate sensitivity to the instantaneous $CO_2$ forcing at the TOA, which preserves the net energy perturbation integrated over all layers of the atmosphere–surface column, yields the prediction of global-mean surface warming in response to $CO_2$ forcing.

For the ERA5 1980–2000 global-mean climate state, the theoretically derived climate sensitivity is 1.15 K $(W\ m^{-2})^{-1}$, corresponding to a total climate feedback parameter of −0.87 $W\ m^{-2}\ K^{-1}$. Applied to the TOA $CO_2$ forcing, it predicts a global warming of 0.44 K from 1980–2000 to 2000–2020, compared with the observed warming of 0.41 K. The diagnosed climate sensitivity using the observed warming and TOA $CO_2$ forcing is 1.06 K $(W\ m^{-2})^{-1}$, which is close to the theoretically derived value. The corresponding equilibrium warming predicted for $CO_2$ doubling and quadrupling is 3.02 K and 5.83 K, respectively, with the predicted warming for $CO_2$ doubling remarkably consistent with the original Charney estimate near 3 °C. We further validate the use of the TOA forcing by explicitly taking into account the vertical profile of the $CO_2$ forcing, which reduces the energy reaching the surface by about 10% in the global mean. The resulting predicted warming is 0.39 K from 1980–2000 to 2000–2020, and 2.73 K and 5.33 K for $CO_2$ doubling and quadrupling, respectively, confirming that applying the climate sensitivity to the TOA forcing provides a good approximation, within about 10%, for the global-mean warming.

The present framework turns climate sensitivity from a quantity diagnosed after the climate response has occurred into one that can be derived from the climate mean state before the response is known. The theoretically derived climate sensitivity can therefore be used to predict global surface temperature responses to general external forcing scenarios at the TOA, including different greenhouse-gas forcing pathways and geoengineering perturbations. When evaluated at each grid point with the local climate mean state and the local vertical profile of the forcing, the same framework also predicts the spatial pattern of global warming, including Arctic amplification, as demonstrated in Cai et al. (2025). Its application to the climate mean states of climate models, which offers a way to constrain the uncertainty in their warming projections, will be reported in a separate study.

**Data availability.** The ERA5 reanalysis is available at https://doi.org/10.5065/P8GT-0R61 (Hersbach et al. 2020). Model codes and data analysis codes will be made available upon reasonable request to enable reproduction of the results.

**Acknowledgements.** X.H. was supported by the National Natural Science Foundation of China (42222502 and 42075028). Z.T. was supported by the National Natural Science Foundation of China (42588101).

**Author contributions.** M.C. and Z.T. initiated and designed the research; M.C. and J.S. performed the research; X.H. contributed source codes and downloaded the data; all authors contributed to discussions and writing.

**Competing interests.** The authors declare no competing interests.

**List of Tables and Figures**

**Table 1.** Climate sensitivity and climate feedback parameter derived from the ERA5 1980–2000 global-mean state.

**Table 2.** $CO_2$ forcing, climate sensitivity, and predicted global-mean surface warming relative to the 1980–2000 state (352.2 ppm). Forcings are in W m$^{-2}$, climate sensitivities in K (W m$^{-2}$)$^{-1}$, and warming in K. $\Delta F^{(trop)}$ is evaluated at 100 hPa.

**Fig. 1.** Radiative coupling matrices for temperature feedback circuits. (a) Planck feedback matrix, (b) matrix $\mathbf{A}$, (c) matrix $\mathbf{U}$, and (d) diagonal matrix of intrinsic circuit gains, $\boldsymbol{G} = \mathrm{diag}(g_i)$. The diagonal elements of both $\mathbf{A}$ and $\mathbf{U}$ are equal to unity, while the off-diagonal elements are positive and less than unity. The diagonal elements of $\boldsymbol{G}$ are positive and less than unity, while all off-diagonal elements are zero. The matrices are computed from the ERA5 1980–2000 global-mean atmosphere–surface column state consisting of 37 layers.

**Fig. 2.** Schematic illustration of the temperature feedback circuits. (a) Source-layer-centered circuit and (b) non-source-layer-centered circuit. The blue circles represent the source layer $i$, where the input energy perturbation $Q_i$ is imposed, while the orange circles represent non-source layers. For the source-layer-centered circuit, the red arrow denotes the unbalanced energy entering layer $i$ prior to the $n$-th feedback cycle, $(g_i)^{n-1}Q_i$, where $g_i = \mathbf{E}_i \cdot \mathbf{A}_i = \sum_{j \neq i} A_{j,i}\, U_{i,j}$, corresponding to the intrinsic circuit gain of layer $i$. The upper pathway represents LW radiation emitted from the source layer $i$ and absorbed by the non-source layers. During the first feedback cycle, the energy absorbed by the non-source layers within this circuit is $\mathbf{A}_i Q_i$, where $\mathbf{A}_i$ denotes the vertical profile of the fraction of energy absorbed by those layers. The lower pathway represents LW radiation emitted from the warmed non-source layers and absorbed by layer $i$. The blue arrow denotes LW radiation emitted to space during the $n$-th cycle. Non-source-layer-centered circuits operate in a similar manner, with each non-source layer $k$ having its own intrinsic circuit gain $g_k = \mathbf{E}_k \cdot \mathbf{A}_k = \sum_{j \neq k} A_{j,k}\, U_{k,j}$, but with an initial energy perturbation, $U_{k,i}Q_i$, that is gained internally from the source layer.

**Fig. 3.** Vertical profiles of surface energy associated with a unit TOA forcing imposed at individual layers of the global–mean atmosphere–surface column. (a) Fraction of the energy imposed at each layer that enters the surface layer prior to amplification by the surface-centered temperature

feedback circuit, given by the surface row of matrix $\mathbf{U}$, $U_{L+1,l}$ (Fig. 1c). (b) Corresponding surface energy after amplification by the surface-centered temperature feedback circuit from the same unit TOA forcing applied at each layer.

**Fig. 4.** Spatial patterns of longwave absorbers and the strength of temperature feedback circuits. (a) Latitudinal profiles of column-integrated water vapor (red; kg m$^{-2}$) and clouds (blue; scaled by a factor of 300 for visualization). Latitudinal variations of (b) $\mathrm{A}_{\mathrm{T_S}}^{(\mathrm{TOA})} = 1/(1 - g_{surface})$, corresponding to the temperature-feedback amplification of the fraction of the TOA forcing that reaches the surface, (c) surface column of matrix **A**, (d) surface row of matrix $\mathbf{U}$, (e) surface column of matrix $\mathbf{U}$, and (f) intrinsic circuit gains of temperature feedback. Note that the color scale in (c) is logarithmic. The results are derived from the ERA5 1980–2000 zonal-mean climatological state. The coordinate of pressure thickness above the ground, given by $(p_s - p)$ in units of hPa, where $p_s$ is the surface pressure and $p$ is the pressure at each level, is used so that zonal averaging is performed at the same level relative to the ground, thereby accurately preserving the vertical structure of longwave radiative coupling within the atmosphere–surface column.

**Fig. 5.** Vertical profiles of global-mean cumulative $CO_2$ radiative forcing from the surface to pressure level $p$. (a) Forcing associated with the increase in $CO_2$ from 352.2 ppm in 1980–2000 to 385 ppm in 2000–2020; (b) forcing for a doubling of $CO_2$ ($2 \times 352.2$ ppm); and (c) forcing for a quadrupling of $CO_2$ ($4 \times 352.2$ ppm). All forcings are calculated from the ERA5 1980–2000 global-mean atmosphere–surface column with a baseline $CO_2$ concentration of 352.2 ppm. The magenta dotted line in each panel indicates the instantaneous $CO_2$ forcing at the TOA, and the red dotted line indicates the instantaneous $CO_2$ forcing at 100 hPa near the tropopause.